\documentstyle[preprint,aps,floats]{revtex}
\tightenlines

\input{epsf.tex}
\input{psfig.tex}

\begin{document}
\def\ba{\begin{eqnarray}}
\def\ea{\end{eqnarray}}
\def\be{\begin{equation}}
\def\ee{\end{equation}}
\def\tr{{\rm tr}}
\def\gtorder{\mathrel{\raise.3ex\hbox{$>$}\mkern-14mu
             \lower0.6ex\hbox{$\sim$}}}
\def\ltorder{\mathrel{\raise.3ex\hbox{$<$}\mkern-14mu
             \lower0.6ex\hbox{$\sim$}}}

\title{Separation Distribution of Vacuum Bubbles in de Sitter Space}

\author{
Carla Carvalho\thanks{E-mail: C.S.N.D.Carvalho@damtp.cam.ac.uk}
and Martin Bucher\thanks{E-mail: M.A.Bucher@damtp.cam.ac.uk}\\
DAMTP, Centre for Mathematical Sciences, University of Cambridge\\
Wilberforce Road, Cambridge CB3 0WA, United Kingdom}

\date{19 July 2002}

\maketitle
\begin{abstract}%
We compute the probability distribution of the invariant
separation between nucleation centers of colliding true vacuum
bubbles arising from the decay of a false de Sitter space vacuum. 
We find that even in the limit of a very small nucleation rate
$(\Gamma H^{-5}\ll 1)$ the production of widely separated bubble pairs
is suppressed. This distribution is of particular relevance for the 
recently proposed ``colliding bubble braneworld'' scenario, in which 
the value of $\Omega _k$ (the contribution of negative spatial
curvature to the cosmological density parameter) is determined
by the invariant separation of the colliding bubble pair. We also
consider the probability of a collision with a `third' bubble.
\end{abstract}

\section{Introduction}

In Guth's original proposal for old inflation \cite{guth}, a false
vacuum with the geometry of de Sitter space decays through
the nucleation as the result of quantum tunnelling of critical bubbles
filled with the true vacuum, presumably Minkowski space or very
nearly so. The dynamics of false vacuum decay, which had been
elucidated by a number of authors \cite{coleman}, may be characterized
by the dimensionless parameter $\Gamma /H^{d+1}$ (where $H$
is the Hubble constant in the false vacuum and $d$ is the 
total number of spatial dimensions, here $d=3$). The 
original `old' inflation envisaged by Guth was unsuccessful
because of the presence of two conflicting requirements. 
On the one hand, for inflation to solve the horizon problem, 
to smooth out whatever inhomogeneities may have existed initially,
and to sweep away the primordial monopoles, $\Gamma /H^{d+1}$
must be small. Yet for inflation to end through the percolation of
a large number of small bubbles, as Guth had originally
envisaged, $\Gamma /H^{d+1}$ had to be large. Although this
potential conflict is noted in Guth's first paper, it was
not until the more detailed work of Guth and Weinberg\cite{gw} that
this incompatibility was quantitatively and definitively
established. 

Although this deficiency of `old inflation' was remedied in the 
work of Linde\cite{linde} and of Albrecht and Steinhardt \cite{steinhardt} in 
`new' and `chaotic' inflation by dispensing with bubbles
altogether, with inflation ending through the `slow roll'
of a scalar field along a relatively flat potential, 
the idea of inflation with bubbles re-emerges in `open 
inflation' \cite{gott}, where re-heating occurs through the evolution
of the inflaton field within the bubble interior. In this
case $\Gamma /H^{d+1}$ is so small that ony a single bubble
is contemplated. 

More recently, the collision of an isolated pair of true vacuum
bubbles filled with (4+1)-dimensional anti de Sitter space 
expanding within a false vacuum of (4+1)-dimensional de Sitter
space has been proposed as a mechanism for setting up a 
Randall-Sundrum-like \cite{rs} cosmology \cite{langlois} with well defined
initial conditions \cite{bucher}. A successful braneworld cosmology
must resolve the (4+1)-dimensional smoothness problem 
\cite{trodden}---that is, explain why the bulk geometry (and any 
additional fields in the bulk as well) respect the same 
three-dimensional spatial homogeneity and isotropy as the 
universe on the brane. It does not suffice to postulate
an epoch of quasi-exponential expansion on the brane itself, 
because after such an inflationary epoch ends, through their
gravitational coupling, imperfections in the bulk will
inevitably induce inhomogeneities on the brane. Because 
anti de Sitter space, quite in contrast to de Sitter space,
does not lose its hair---the amplitudes of perturbations
remain constant rather than decay---some other mechanism
to explain the spatial homogeneity and isotropy of the bulk, 
and to establish a preferred initial quantum state is required. 

In the colliding bubble braneworld scenario, an initial
(4+1)-dimensional de Sitter epoch sweeps away whatever
inhomogeneities may have previously existed. After a
modest number of expansion times, the quantum fluctuations
about the classical vacuum rapidly approach the Bunch-Davies
vacuum \cite{bd}. Classically, the nucleation of a single true vacuum
bubble breaks the $SO(5,1)$ symmetry of $dS^5$ to $SO(4,1),$
the subgroup of transformations that leave the nucleation
center of the bubble invariant. For the case of two bubbles,
this residual symmetry is further broken to $SO(3,1),$\cite{hawking}
the subgroup of those transformations that leave both 
nucleation centers invariant. The three-dimensional, completely
spatial surface of bubble collision, from which our local brane
endowed with a modified (3+1)-dimensional FLRW cosmology arises,
is endowed with the geometry of $H^3,$ of constant negative
spatial curvature. 

Owing to the negative spatial curvature of this geometry, the eventual
fate of such a universe (in the absence of a non-vanishing cosmological
constant or something similar) is that of an empty expanding universe
whose expansion is predominantly driven by its spatial curvature, as in a Milne 
universe. However, because of the exponential suppression of the 
nucleation of bubbles, this dreary fate may lie far in our future. 
It is not difficult to envisage bubble separations so large that
today the spatial curvature is still negligible. 
$\Omega _0,$ the cosmological density parameter today (when $T_{rad}=2.75K$),
is a monotonically increasing function of the bubble 
pair separation.  The farther apart the bubbles, the larger
$\rho _{coll}$ and the smaller the spatial curvature on the surface 
of collision. The radius of spatial curvature of the surface of
collision is given by\cite{bb}
\be
R={H_{dS}}^{-1}\tan [\sigma /2],
\ee
which diverges as $\sigma \to \pi.$ 
Here $\sigma $ is the spatial geodesic separation between the nucleation centers
in units where $H^{-1}=1,$ so that $\sigma =\pi $ corresponds to bubble separated
exactly at the threshold of never striking each other. 
We observe that even though
$\sigma $ for colliding bubbles
is bounded from above by a finite limit, the surface of 
collision can be arbitrarily close to flat. The gamma factor
$\gamma =1/\sqrt{1-(v/c)^2}$ of the colliding bubbles in the 
local center-of-mass frame of the collision, which is equal to
\be
\gamma _{max}={\sin [\sigma /2] \over \sin [r_c]}.
\ee
where $r_c$ is the radius of the critical bubble
(in units in which ${H_{dS}}^{-1}=1),$ 
however, is limited by a finite upper bound.\footnote{Note that
the notation here differs from that of ref.~\cite{bb}.} 
Both effects drive $\Omega _0$ toward unity as $\sigma \to \pi .$ The precise
relationship between $\sigma $ and $\Omega _0$ depends
on the details of the equation of state after the collision, 
which is unknown and model dependent. 

In the colliding bubble scenario the value of $\Omega $ today
is a random variable dependent on the interbubble separation
which is stochastically determined. The object of this article
is to study the distribution of the interbubble separations
for the case of bubbles nucleating in de Sitter space. 

\section{Bubble Collision Probabilities}

We now compute the probability distribution for the interbubble
separation of pairs of colliding bubbles. This separation $\sigma $ 
is defined in a coordinate free manner as the length of the (shortest) 
spacelike geodesic connecting the nucleation centers of the two 
colliding bubbles. We adopt units in which $H=1$ and assume a spatially and temporally
uniform nucleation rate $\Gamma ,$ the absence of correlations
of the nucleations of nearby bubbles, and that bubbles cannot
nucleate inside regions already within bubbles. 

First a lightning review of the geometry of de Sitter space. 
Maximally extended (d+1)-dimensional de Sitter 
space $dS^{(d+1)}$ is most readily constructed as the 
embedding of the unit hyperboloid defined by 
\begin{equation}
\label{hyperboloid}
-T^2+{X_0}^2+\ldots +{X_d}^2=1
\end{equation}
in (d+2)-dimensional Minkowski space $M^{(d+2)}$ with the line element
\begin{equation}
\label{met1}
ds^2= -dT^2+d{X_0}^2+\ldots +d{X_d}^2.
\end{equation}
This space in its entirety may be covered by the coordinates 
\ba
 T &=& \sinh [\tau ], \nonumber\\
 X_0 &=& \cosh [\tau ]~n_0,\nonumber\\
 &&\ldots \nonumber\\
 X_d &=& \cosh [\tau ]~n_d,
\ea
where $(n_0,\ldots , n_d)$ are the coordinates of a point on the unit
sphere $S^d$ embedded in $E^{(d+1)},$ with the line element 
\begin{equation}
\label{met2}
ds^2= -d\tau ^2 + \cosh ^2[\tau ]d\Omega_{(d)}^2.
\end{equation}

Any two points with a spacelike separation and the property that their
forward lightcones eventually intersect may be transformed to lie on the
throat of the hyperboloid defined in eqn. (\ref{met1}) (i.e., the sphere
$T=0).$ For the largest such spatial separation, $\sigma =\pi ,$ and 
two points are antipodal. This situation (with our 
idealization of zero critical bubble radius) corresponds to bubbles
which almost but never collide. Larger spatial separations are also
possible, but in this case no spacelike geodesics exist linking the 
two points. 

As a technical device, it is useful to characterize the invariant separation
between two points $P$ and $Q$ in terms of the (d+2)-dimensional 
Minkowski space inner product 
\begin{equation}
I(P,Q)=-T_PT_Q+\sum _{i=0}^d(X_i)_P~(X_i)_Q.
\end{equation}
It follows that
\begin{equation}
I(P,Q)=\cos [\sigma ]
\end{equation}
for spacelike separated points of the first kind and that
\begin{equation}
I(P,Q)<-1
\end{equation}
for those of the latter kind. 
(Timelike separations correspond to $I(P,Q)>+1.$)

\begin{figure}[h]
\vskip 15pt
\begin{picture}(600,150)
\put(200,20){\leavevmode\epsfxsize=2in\epsfbox{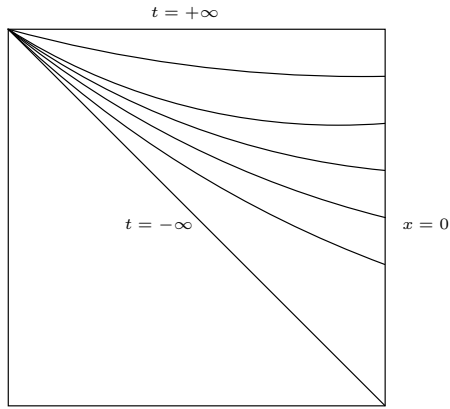}}
\put(350,88){\tiny $x=0$}
\put(245,88){\tiny $t=-\infty $}
\put(255,168){\tiny $t=+\infty $}
\end{picture}
\vskip -10pt
\caption{
The spatially flat coordinate slicing covers precisely half of maximally
extended de Sitter space, as indicated in the Penrose-Carter
conformal diagram above. The two vertical boundaries are identified and 
on the horizontal boundaries the conformal factor diverges. 
}
\label{fig:1}
\end{figure}

In terms of the flat coordinates, which cover precisely half of 
maximally extended de Sitter space, as indicated in Fig.~1,
the line element is
\be
ds^2=-dt^2+e^{2t}\cdot \Bigl[ d{x_1}^2+\ldots +d{x_d}^2\Bigr],
\ee
which after the change of variable to conformal time $\eta =-e^{-t},$
with $-\infty <\eta <0,$ becomes 
\be
ds^2={1\over \eta ^2}\cdot \Bigl[ -d\eta ^2+d{x_1}^2+\ldots +d{x_d}^2\Bigr] .
\ee
The embedding of these coordinates into $M^{(d+1)}$ is given by
\ba
T  &=&\sinh [t]+e^t~{\bf x}^2/2,\nonumber\\
X_0&=&\cosh [t]-e^t~{\bf x}^2/2,\nonumber\\
X_i&=&e^t~x_i,
\ea
where 
${\bf x}\cdot {\bf y}=x_1y_1+\ldots +x_dy_d.$ The invariant 
separation defined above is given by
\be 
I(P,Q)={{\eta _P}^2+{\eta _Q}^2-({\bf x}_P-{\bf x}_Q)^2\over 2\eta _P\eta _Q}.
\ee

We now turn to the issue of initial conditions. It is necessary to specify
an initial surface over which there is everywhere false vacuum.
As has been previously pointed out in the literature\cite{borde}, an unstable de
Sitter vacuum extended infinitely far into the past of maximally extended
de Sitter space does not make any sense. Bubbles formed during the contracting
phase eventually percolate. However, if consideration is restricted to 
a subregion of de Sitter space over which there is only expansion (such as
that covered by the flat coordinates), then, if $\Gamma $ is
sufficiently small, inflation is eternal in the forward time direction
(i.e., the bubbles never percolate), and presumably any conclusions
drawn are insensitive to the character of this initial surface. 

We employ the flat coordinates, assuming a false vacuum everywhere on the 
surface $t=t_-$ and considering only bubbles that nucleate for $t$ in the 
interval $t_-<t<t_+.$ (Eventually we take the double limit
$t_-\to -\infty ,$ $t_+\to +\infty .$) 

\begin{figure}[h]
\vskip 15pt
\begin{picture}(600,150)
\put(150,0){\leavevmode\epsfxsize=3in\epsfbox{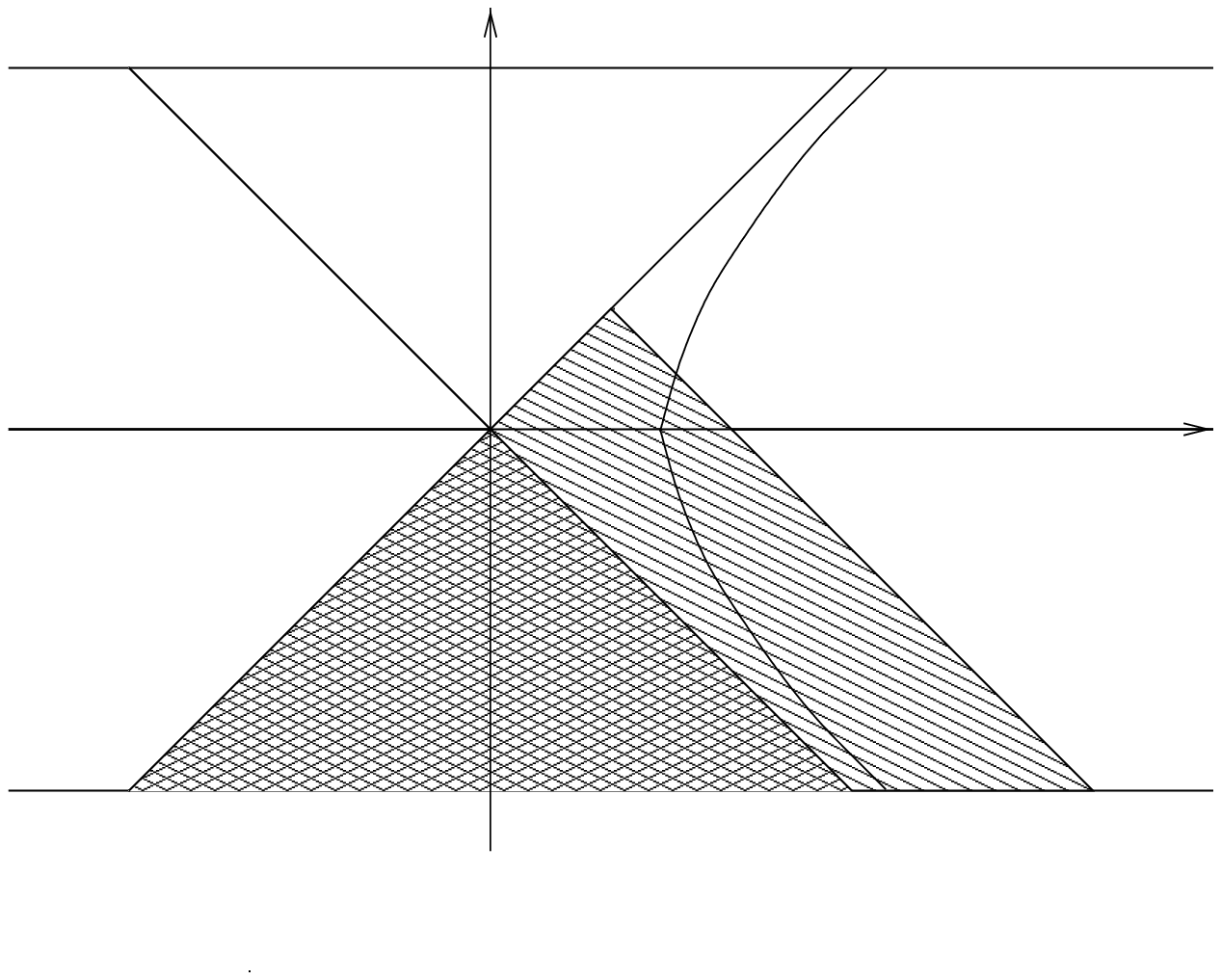}}
\put(290,60){\small $V_{excl}$}
\put(250,60){\small $V_{lc}$}
\put(238,104){\small $P$}
\put(275,108){\small $Q$}
\put(290,128){\small $\Sigma $}
\end{picture}
\vskip -10pt
\caption{
We take as given that a bubble has nucleated at the point $P.$
The surface $\Sigma $ indicates the surface of possible nucleation centers
separated by the same proper distance $\sigma $ from $P.$ For a bubble that nucleates
at $Q$ on $\Sigma $ to collide with the bubble that nucleated at $P$
(rather than with some other intermediate bubble), it is necessary that no bubble nucleated
in the singly shaded region.
}
\label{fig:2}
\end{figure}

The probability density for a bubble to nucleate at time $t$
among all the bubble nucleating in the time interval
$t_-<t<t_+$ is given by the normalized density 
\begin{equation}
\frac{dP_{bn}}{dt~~}(t)=
e^{dt}~e^{-\Gamma\, V_{lc}(t)}\times 
\left[ \int ^{t_{+}}_{t_{-}}dt'~e^{dt'}~e^{-\Gamma\, V_{lc}(t')}\right] ^{-1}
\label{probbn}
\end{equation}
where $V_{lc}$ is the volume of the past light cone of the nucleation
center lying above the initial surface $t=t_-.$  Here we have considered 
an infinitesimally thin tube of constant co-moving d-volume, whose physical
d-volume scales as $e^{dt}.$

Next we consider bubble collisions. At first we restrict ourselves to
the simplest case of bubble in (1+1)-dimensional de Sitter space to
avoid the complications arising from the fact that in higher dimensions
a bubble can collide with many other bubbles. However, for all volumes 
and measures we retain the general formulae, valid for all $d.$ 
In (1+1)-dimensions, each bubble collides with precisely one bubble from 
the left and one bubble from the right. Once the (1+1)-dimensional case 
has been thoroughly examined, we generalize to arbitrary dimension in the small 
$\Gamma $ limit.

We now calculate the conditional probability $dP_{bc}(\sigma \vert t)/d\sigma $
that, given that a first bubble has nucleated at time $t,$ another bubble whose
nucleation center is separated from that of the first bubble by a proper
distance $\sigma $ collides with the first bubble. This probability density
is given by the expression
\begin{equation}
\frac{dP_{bc}}{d\sigma }(\sigma \vert t)= 
\int_\Sigma 
d\Sigma ~ \Gamma ~e^{-\Gamma V_{excl}(\Sigma )} .
\label{probbc}
\end{equation}
The various volumes are indicated in Fig.~2. $P$ is the nucleation center
of the bubble that has by assumption nucleated at time $t.$ The presence of
this bubble implies that no other bubble has nucleated  
in the part of its past lightcone to the future of the initial surface
(i.e., the doubly shaded region of the figure). $\Sigma $ is the locus of all points
a proper distance $\sigma $ to the right of $P$ and $d\Sigma $ is the 
volume element on this surface. $Q$ is a point on this surface, 
a potential nucleation center from a bubble a proper distance
$\sigma $ from $P$ that may collide with the bubble emanating from $P.$
However, for such a collision to occur, it is necessary that no
bubble nucleate in the lightly shaded region, of volume $V_{excl}.$
It is necessary to exclude the case of a bubble nucleating in the 
volume $V_{excl},$ for any such bubble would collide first with
the bubble emanating from $P.$ The probability of no nucleation
in the lightly shaded region is $\exp (-\Gamma V_{excl}),$ hence
the appearance of this factor in eqn.~(\ref{probbc}). If we remove 
the upper cut-off, taking $t_+\to +\infty ,$ the integral of the density 
from $\sigma =0$ to $\sigma =\pi $ becomes equal to unity---in other words,
a collision occurs with unit probability.  

We may combine eqns.~(\ref{probbn}) and (\ref{probbc}) to obtain 
the probability density properly averaged over the nucleation time
for the first bubble $t$ for collisions of bubbles separated by 
a proper distance $\sigma $
\be
\frac{dP_{av}(\sigma )}{d\sigma }  = \int^{t_{+}}_{t_{-}} dt ~
\frac{dP_{bc}}{d\sigma }(\sigma |t)~
\frac{dP_{bn}(t)}{dt}.
\label{prob}
\ee

Particularizing to (1+1) dimensions and replacing the proper time $t$
with the conformal time $\eta ,$ we rewrite eqn.~(\ref{probbc}) as 
\begin{equation}
\frac{dP}{d\cos[\bar \sigma ]}= 
\Gamma \int _{-\infty }^0\frac{d\eta}{\eta^2}\int _0^\infty dx ~
\delta \Bigl( \cos [\sigma ]  -\cos  [\bar \sigma ] \Bigr)  ~e^{-\Gamma V_{excl}}. 
\label{probdS}
\end{equation}
Without loss of generality we may set the coordinates of $P$ to
$\eta _P=-1,$ $x_P=0.$
We eliminate the integration over $x$ by considering 
\be
f(x)=\cos [\sigma ]={\eta ^2+{\eta _P}^2-(x-x_P)^2\over 2\eta ~\eta _P}
={x^2-\eta ^2-1\over 2\eta },
\ee
so that eqn.~(\ref{probdS}) becomes
\ba
&&\Gamma \int _{-\infty }^0
{d\eta \over \eta ^2}
\left| {\partial f\over \partial x}\right| ^{-1}~
e^{-\Gamma V_{excl}\bigl(x(\eta ), ~\eta \bigr) }\nonumber\\
&=&\Gamma \int _{-\infty }^0
{d\eta \over \eta ^2}
\left( {-\eta \over x}\right) 
e^{-\Gamma V_{excl}\bigl(x(\eta ), ~\eta \bigr) }\nonumber\\
&=&\Gamma \int _{-\infty }^0
{d\eta \over \vert \eta \vert }
{1\over \sqrt{1+2\eta \cos [\bar \sigma ]+\eta ^2}}
e^{-\Gamma V_{excl}\bigl(x(\eta ), ~\eta \bigr) }.
\label{MBAA}
\ea

We now turn to the computation of $V_{excl}(x, \eta ).$ As can readily be 
seen from Fig.~2, this volume depends only on the point 
of intersection $(\bar x, \bar \eta )$ of the left-moving null 
geodesic through $(x, \eta )$ with the right-moving null geodesic
emanating from $P.$ It follows that
\be
\bar \eta =\frac{1}{2}\Bigl[ \eta +x -1 \Bigr].
\ee
We compute 
\be
V_{excl}(\bar \eta )=\int _{-\infty }^{-1}{d\eta \over \eta ^2}~
2(1-\vert \bar \eta \vert )+
\int _{-1}^{-\vert \bar \eta \vert }{d\eta \over \eta ^2}~
2(-\vert \bar \eta \vert -\eta )
=2\ln \left( 1\over \vert \bar \eta \vert \right) .
\ee
Consequently, eqn.~(\ref{MBAA}) becomes
\ba
{dP\over d\cos [\bar \sigma ]}&=&
\Gamma \int _{-\infty }^0{d\eta \over |\eta |}
{1\over \sqrt{1+2\eta \cos [\bar \sigma ]+\eta ^2}}
\exp \left[ +2\Gamma \log \left( {(1-\eta -\sqrt{1+2\eta \cos [\bar \sigma ]+\eta^2}\over 2}
\right) \right]\nonumber\\
&=&\Gamma \int _0^{+\infty }{d\eta \over \eta }
{1\over \sqrt{1-2\eta \cos [\bar \sigma ]+\eta ^2}}
\left[ {1+\eta -\sqrt{1-2\eta \cos [\bar \sigma ]+
\eta^2}\over 2}\right] ^{2\Gamma }.
\ea
For small $\Gamma ,$ the dominant contribution to the integral lies
in the neighborhood of $\eta =0.$ Therefore, we may approximate
the above expression as
\ba
&&\Gamma \int _0^{+\infty }{d\eta \over \eta }
{1\over \sqrt{1-2\eta \cos [\bar \sigma ]+\eta ^2}}
\left[ {\eta (1+\cos [\bar \sigma ])\over 2}\right] ^{2\Gamma }
\nonumber\\
&\approx &\Gamma \left( {1+\cos [\bar \sigma ]\over 2}\right) ^{2\Gamma }
\int _0^{1}d\eta ~\eta ^{(2\Gamma -1)}\nonumber\\
&\approx &\frac{1}{2}
\left( {1+\cos [\bar \sigma ]\over 2}\right) ^{2\Gamma }\approx 
\frac{1}{2}.
\ea
In other words, for $\Gamma \to 0+,$ we obtain a uniform distribution in 
$\cos \sigma ,$ or 
\be
{dP\over d\sigma }={\sin [\sigma ]\over 2}.
\ee

We now generalize to $d>1.$ Since any given bubble suffers an indeterminate
number of collisions with other bubbles, it is not at the outset clear 
how to weight the various collisions to obtain a sensible distribution
for $dP/d\sigma .$ We first define the distribution in a rather artificial
way, later showing that the result obtained is more generally valid in 
the small $\Gamma $ limit. 

We start by postulating as given the nucleation of a bubble, taken without
loss of generality to be situated at $\eta =-1,$ ${\bf x}=0,$ and compute
the distribution for the first other bubble to collide with this bubble. (This
may seem a bit artificial because a preferred time direction has been
singled out.) Defined in this way, eqn.~(19) becomes
\ba 
V_{excl}(\vert \bar \eta \vert )={\Omega _{d-1}\over d}
\Biggl[ &&
\int _{\vert \bar \eta \vert }^1{d\eta \over \eta _{d+1}}~~
\Bigl[ (\eta +1-2\vert \bar \eta \vert)^d- (1-\eta )^d\Bigr]
\nonumber\\
&+&
\int _1^{+\infty }{d\eta \over \eta _{d+1}}
\Bigl[ (\eta +1-2\vert \bar \eta \vert)^d- (\eta -1)^d\Bigr]
\Biggr]
\ea
where $\Omega _{d-1}$ is the area of the (d-1)-unit sphere. In the small
$\Gamma $ limit, we are only interested in the most singular part of
$V_{excl}$ as $\vert \bar \eta \vert \to 0+.$ In the limit, the second
integral in the expression above is finite, and consequently can be 
ignored. The first integral is dominated by the contribution in the 
neighborhood of the lower limit of integration, which may be approximated
as 
\be
V_{excl}\approx \Omega _{d-1}\int _{\vert \bar \eta \vert }^1
{d\eta \over \eta ^{d+1}}
2(\eta -\vert \bar \eta \vert )\approx 
{2\Omega _{d-1}\vert \bar \eta \vert ^{-d+1}\over d(d-1)}.
\ee
We obtain (setting $\bar \Gamma =\Gamma \Omega _{(d-1)}$)
\ba
{dP(\sigma )\over d\cos [\sigma ]}
&=&\Gamma \Omega _{d-1}
\int _0^{+\infty }{d\eta \over \eta ^d}
{1\over \sqrt{1-2\eta \cos [\sigma ]+\eta ^2}}
\exp \left[ {-2\Gamma \Omega _{d-1} \vert \bar \eta \vert ^{-d+1}\over d(d-1)}\right]
\nonumber\\
&\approx & \bar \Gamma 
\int _0^{+\infty }{d\eta \over \eta ^d}~~
\exp \left[ {-2\bar \Gamma \over d(d-1)}
\left({\eta (1+\cos [\sigma ])\over 2}\right)^{-d+1} \right]
\nonumber\\
&\approx & {\bar \Gamma\over (d-1)}
\int _0^{+\infty }dz~
\exp \left[ {-2\bar \Gamma \over d(d-1)}
\left({(1+\cos [\sigma ])\over 2}\right)^{-d+1} z\right]
\nonumber\\
&=&{d\over 2}\left( {1+\cos [\sigma ]\over 2}\right) ^{d-1}=
{d\over 2}
\left( \cos ^2 \left[{\sigma \over 2}\right] \right) ^{d-1},
\ea
so that
\be 
{dP(\sigma )\over d\sigma }={d\over 2}
\sin\sigma 
\left( {1+\cos [\sigma] \over 2}\right) ^{d-1}.
\label{eee}
\ee
We observe a power low suppression for large bubbles.

We now explain why the distribution above is more broadly valid,
when multiple collisions have been taken into account. As before, we 
consider a bubble whose nucleation center has been placed at
$\eta =-1,$ ${\bf x}=0.$ We consider an infinitesimal region on 
the surface of the bubble from $(\bar \eta , 1+\bar \eta )$
and $(\bar \eta +d\bar \eta, 1+\bar \eta +d\bar \eta )$ and some
interval of solid angle as well, and consider, given that some
exterior bubble first strikes this bubble in that patch, what
is the distribution of $\sigma .$ Computing this distribution
involves extending the patch radially outward and into the past
along the relevant null geodesics. Let $\xi $ range from $0$
to $+\infty .$ Then 
\ba 
\eta &=& \bar \eta -\xi,\nonumber\\
r    &=& 1+ \bar \eta +\xi .
\ea
It follows that 
\be
\cos [\sigma ]={r^2-\eta ^2-1\over 2\eta }=
{\bar \eta +\xi \over \bar \eta -\xi }.
\ee
Therefore, $\xi =-\bar \eta \tan ^2 [\sigma /2].$
For small $\Gamma ,$ we may take $\bar \eta $ to be very close to
zero, because all but a few bubbles nucleate very long after the 
nucleation of the first bubble.  Using the fact 
that
\be
{dP\over d\cos [\sigma ]}\sim {dV\over d\cos [\sigma ]}
\ee
we obtain
\be
dV\sim d\xi \left( {r^{d-1}\over \eta ^{d+1}}\right) \approx 
{d\xi \over \eta ^{d+1}}={d\xi \over (\bar \eta -\xi )^{d+1}}
\sim \left( \cos ^2 \left[{\sigma \over 2}\right] \right)^ {d-1}d(\cos [\sigma] )
\ee
confirming the result in eqn.~(\ref{eee}).
By means of this last calculation, all bubble collisions for which a 
portion of the bubble collision surface is tangent to one of the 
$t=(constant)$ surfaces are counted and given equal weight. 

\section{Collisions with a Third Bubble}

In the colliding bubble braneworld scenario the universe that arises from the 
collision of two bubbles has a hyperbolic geometry
that would be infinite in extent in the absence of any collision 
with a third bubble. However, it is inevitable that sooner or later
collisions with third bubbles occur, and such collisions 
truncate the extent of this hyperbolic universe. Physically, the 
relevant question is whether such a cosmological scenario is likely
to contain a hyperbolic patch large enough to contain the 
entire universe observable to us today, for the consequences of
a collision within our past lightcone with a third bubble would
hardly be subtle. At a very minimum, such a collision would create an
$O(1)$ perturbation in $\Delta T/T$ of the CMB anisotropy. More dire
consequences are also possible. 

Concretely, to obtain an estimate of the likelihood of a 
collision with a third bubble within our past lightcone, we consider 
how the infinite $H^{(d-1)}$ surface of collision of the two bubbles is
truncated by collisions with other bubbles. We take as given the initial
collision of the two bubbles, assuming at least that the point $\xi =0$ 
on $H^{(d-1)}$ has been spared from a collision with a third bubble. Here
$ds^2=R^2\Bigl( d\xi ^2+\sinh ^2[\xi ]d\Omega _{(d-1)}^2\Bigr) $ is
the metric on $H^{(d-1)}$.) We calculate the probability distribution for the 
radius $\xi $ of the largest sphere $B_{\xi }= 
\{ (\bar \xi ,\theta ,\phi ) \vert \bar \xi < \xi \}\subset H^{(d-1)}$ such 
that no bubble has nucleated within the past lightcone of $B_{\xi }.$
As before we employ
the coordinates 
$ds^2=(1/\eta ^2)(-d\eta ^2+d{\bf x}^2)$
for the larger $(d+1)$-dimensional de Sitter space 
into which $H^{(d-1)}$ is embedded. 
If the integration over the volume contained within a past 
lightcone of a spacetime point is taken all the way back
to $\eta =-\infty ,$ the resulting integral diverges logarithmically 
as $\ln \Bigl( \vert \eta _{min}\vert \Bigl) .$
However, since by assumption no bubble has nucleated in the past
lightcone of the origin $O$ at $\xi =0,$ we are interested in the volume 
$V_{\xi }$ that lies between the past lightcone of $B_{\xi }$ 
and the past lightcone of the origin $O,$ and the integral defining
$V_{\xi }$ is convergent. One obtains
\ba
\frac{dP}{d\xi }=\Gamma \frac{dV}{d\xi }\exp 
\Bigl[ -\Gamma V(\xi )\Bigr] .
\label{eqn:pd}
\ea
For a typical value of $\xi$, $\Gamma V(\xi) \approx 1$.

We first consider the simplest case with $(d=2),$ later 
straightforwardly generalizing to $d>2.$ We define the hyperboloid
$H^{(d-1)}$ as the locus of points a proper time $\tau $ to the future
of its focus, where proper time is calculated using the de Sitter
metric and $\sinh [\tau ] =R$. Without loss of generality we fix the
focus to the point $\eta =-1, 
{\bf x}=0.$ From the relation for the invariant separation in de Sitter
space, it follows that the hyperboloid is defined by the expression
\ba
\frac{1+\eta ^2-{\bf x}^2}{-2\eta }=
\frac{1+\bar \eta ^2}{-2\bar \eta }=
\cosh [\tau ]=\sqrt{1+R^2}. 
\ea
Here $(\bar \eta, {\bf 0})$ is the spacetime position vector of the
origin of the hyperboloid $O$.
It follows that $R=(1-\bar \eta ^2)/(-2\bar \eta ).$ 
We find it convenient 
to parameterize the hyperboloid according to
\ba 
\eta (\theta )&=& \hbox to .99in{$-\sqrt{1+R^2}+$}R\sec [\theta ],\nonumber\\
   r (\theta )&=&  \hbox to .99in{}R\tan [\theta ] ,
\ea
where $0\le \theta <\theta _{max}$, such that $\tan [\theta
_{max}]=1/R.$ As $R \to \infty$, $\theta_{max} \to 0$.

In terms of the distance in the larger de Sitter
space, the line element along the hyperboloid gives
\ba
\frac{ds}{d\theta }=\frac{R\sec [\theta ]}%
{\sqrt{1+R^2}-R\sec [\theta ]},
\ea
or in terms of the natural dimensionless distance on $H^{(d-1)}$
\ba
\frac{d\xi }{d\theta }=\frac{\sec [\theta ]}%
{\sqrt{1+R^2}-R\sec [\theta ]}.
\label{eqn:txt}
\ea

To compute $dV/d\xi $ in eqn.~(\ref{eqn:pd}) at a certain value of 
$\xi ,$ we use the relation
\ba
\frac{dV}{d\xi}=
\frac{d\theta }{d\xi}
\frac{dV}{d\theta }
\ea
and compute the infinitesimal volume $(dV/d\theta )d\theta $ swept
out by the displacement $d\theta $ as the past lightcone at $\theta $
is displaced to $(\theta +d\theta ).$ In the underlying Minkowski
space, with metric $ds^2=-d\eta ^2+d{\bf x}^2,$ the cone at 
$P(\theta )=(\eta (\theta ),r(\theta ),0)$ suffers a rigid displacement by
\ba
\frac{~\partial }{\partial \theta }=R\sec [\theta ]\cdot \Biggl[
\tan [\theta ]\frac{~\partial }{\partial \eta }
+\sec [\theta ]\frac{~\partial }{\partial r}
\Biggr] .
\ea

We parameterize the surface of the cone using $w,$ $w_{min}\le w\le +\infty ,$
and $\psi ,$ $ -\pi <\psi <+\pi ,$ where $w_{min}=-\eta (\theta ),$
\ba
\eta &=& -w,\nonumber\\
r &=& (w-w_{min})\cos [\psi ],\nonumber\\
x &=& (w-w_{min})\sin [\psi ]. 
\ea

The volume swept out by a surface element $(dw)(d\psi )$ is given by the 
determinant
\ba
\frac{dV~~~~}{d\psi dw~d\theta }&=&
R\sec [\theta ] (w-w_{min})
\left|
\matrix{
\tan [\theta ]& \sec [\theta ]& 0\cr 
-1& \cos [\psi ]& \sin [\psi ]\cr 
0& -\sin [\psi ]& \cos [\psi ]\cr 
}
\right| \nonumber\\
&=&
R\sec [\theta ] (w-w_{min})
\Bigl[ \tan [\theta ]+\sec [\theta ]\cos [\psi ]\Bigr] \nonumber\\
&=&
R\sec ^2[\theta ] (w-w_{min})
\Bigl[ \cos [\psi ]-\cos [\pi /2+\theta ]\Bigr] .
\ea
Since we are interested only in the volume swept out in the 
forward direction, we restrict the integration to the subregion over 
which this determinant is positive---in other words, over the range
$\vert \psi \vert <\pi /2+\theta .$ Including the conformal 
factor $w^{-3}$ to reflect de Sitter rather than Minkowski volume,
we obtain 
\ba 
\frac{dV}{d\theta }&=&2R\sec ^2[\theta ]\int _{\vert \eta (\theta )\vert }^{+\infty }
\frac{dw}{w^3}(w-\vert \eta (\theta )\vert )
\int _0^{\pi /2+\theta }d\psi \Bigl[
\cos [\psi ]-\cos [\pi /2+\theta ]
\Bigr]  \nonumber\\
&=&\frac{R\sec ^2[\theta ]}{\sqrt{1+R^2}-R\sec [\theta ]}
\left[
\cos [\theta ]+\left( \frac{\pi}{2}+\theta \right)
\sin [\theta ] \right] , 
\label{eqn:rsta}
\ea 
which in light of eqn.~(\ref{eqn:txt}) becomes 
\ba
\frac{dV}{d\xi }=R\left[ 1 +\left( \frac{\pi}{2}+\theta \right)
\tan [\theta ] \right] . 
\ea
We are concerned with the limit $\Gamma\ll 1$, in which the values
of $\theta$ of interest lie very near  $\theta_{max}$. In this limit, $r \to 1$ and
$\tan \theta \to 1/R.$ For large $R,$ $\theta _{max}$ is small, and $dV/d\xi \approx R,$
giving $V(\xi )\approx R\xi .$ Since $\Gamma $ is exponentially small, the requirement 
that $\Gamma V(\langle  \xi \rangle )=O(1),$ where here $\langle \xi \rangle $ is the 
expectation value of $\xi ,$ implies that $\langle \xi \rangle $ is exponentially large. 

To generalize to higher dimensions, we must modify eqn.~(\ref{eqn:rsta}) by changing the 
conformal factor from $w^{-3}$ to $w^{-(d+1)}$ and replacing the factor of 
$\Omega _0=2,$ with $\Omega _{(d-2)}$ where 
$\Omega _1=2\pi ,$ $\Omega _2=4\pi ,$ $\ldots .$
Moreover, $dr$ becomes $r^{d-1}dr,$ where now the position 
of the origin for $r$ does matter. With these modifications,
eqn.~(\ref{eqn:rsta}) is now
\ba
\frac{dV}{d\theta }&=&\Omega _{(n-2)}R\sec ^2[\theta ]
\int _{\vert \eta (\theta )\vert }^{+\infty }
\frac{dw}{w^{d+1}}(w-\vert \eta (\theta )\vert )\nonumber\\
&&\times \int _0^{\pi /2+\theta }d\psi \Bigl[
\cos [\psi ]-\cos [\pi /2+\theta ]
\Bigr]  \Bigl[ r(\theta )+\Bigl( w-\eta (\theta )\Bigr) \cos [\psi
]\Bigr] ^{d-2}\nonumber \\
&=&\Omega _{(d-2)}R\sec ^2[\theta ]\frac{1}{\vert \eta(\theta)\vert}
\int _{1}^{+\infty }
\frac{du}{u^{d+1}}(u-1)\nonumber\\
&&\times \int _0^{\pi /2+\theta }d\psi \Bigl[
\cos [\psi ]-\cos [\pi /2+\theta ]
\Bigr]  \left[\frac{r(\theta )}{\vert \eta(\theta)\vert}+(u-1) \cos [\psi ]\right] ^{d-2}
\label{eqn:rstb}
\ea
where $u=w/\vert \eta(\eta)\vert$.

For the dimension of greatest interest ($d=4$), it is not possible to evaluate
\be
V(\theta)=\int^{\theta}_0 d\theta \frac{dV}{d\theta} 
\ee
with the integrand defined in eqn.~(\ref{eqn:rstb}). However, for
$\Gamma\ll 1$, we are interested in $\theta$ very near $\theta_{max}$,
so that $r(\theta)/ \vert \eta(\theta) \vert \gg 1$. Hence, we may
drop all but the leading term in powers of $r(\theta)/ \vert
\eta(\theta) \vert$ in eqn.~(\ref{eqn:rstb}), which will then reduce to
\be
\frac{dV}{d\theta} \approx R \sec^2 [\theta] \frac{r^2(\theta)}{\vert
\eta(\theta)\vert ^3} 
\left[
\cos [\theta ]+\left( \frac{\pi}{2}+\theta \right)
\sin [\theta ] \right].
\label{eqn:rstb1}
\ee
As $\theta \to \theta _{max},$ $r\to 1$ and 
\be
\frac{1}{\vert \eta(\theta) \vert} = \frac{\cos[\theta]
\sin[\theta_{max}]}{\cos[\theta]-\cos[\theta_{max}]}
\approx \frac{\cos[\theta_{max}]}{(\theta_{max}-\theta)},
\label{eqn:eta_apprx}
\ee
so that 
\ba
V &\approx& \left(R+\frac{\pi}{2}+\theta_{max}\right) \cos ^2[\theta_{max}] \int d\theta
\frac{1}{(\theta_{max}-\theta)^3} \nonumber\\
&\approx&  \left(R+\frac{\pi}{2}+\theta_{max}\right)
\left(\frac{R^2}{1+R^2}\right) 
\left[\frac{1}{2 (\theta_{max}-\theta)^2} + O\left(\frac{1}{ (\theta_{max}-\theta)}\right)\right].
\label{eqn:rstb2}
\ea
{}From eqn.~(\ref{eqn:txt}) it follows that 
\ba
\frac{d\xi }{d\theta }\approx \frac{1}{(\theta _{max}-\theta )}.
\ea
Therefore,
\ba
\xi \approx -
\ln [\theta _{max}-\theta ].
\ea
and 
\ba 
(\theta _{max}-\theta )\approx \exp \left[ -\xi \right] 
\ea
and 
\ba 
V\approx \frac{1}{2}(R+\frac{\pi }{2}+\theta _{max})\left( \frac{R^2}{1+R^2}\right) 
\exp \left[ +2\xi \right] .
\ea 
Given that \cite{bucher}, \cite{bb} 
\ba 
\Gamma \approx \exp [ -A(m_4\ell )^2]
\ea
where $A=O(1),$ 
we find that for $(m_4\ell )\gg 1,$
\ba
\left\langle \xi \right\rangle \approx \frac{A}{2} (m_4\ell )^2
\ea
where we have suppressed all logarithmic corrections. This result implies that
for large $(m_4\ell ),$ the $SO(3,1)$ symmetric regions where the bubbles collide 
are composed of patches of a size almost always containing several
curvature lengths, suggests that one is unlikely to observe a collision 
with a third bubble.

\section{Discussion}

The object of this study was twofold: first, to determine whether in the limit
$\Gamma \to 0$ it is possible to obtain, with a non-negligible probability,
bubble
separations very near but just short of the maximal separation $\sigma =\pi ,$ 
where the bubbles just barely strike each other; and, second, to determine 
whether collisions with a third bubble constrain or rule out the colliding 
bubble braneworld scenario. 

With respect to the first question, we found that as $\Gamma \to 0+,$ 
the probability distribution $dP/d\sigma $ approaches a limit independent 
of $\Gamma ,$ given in eqn.~(\ref{eee}), having a power law 
suppression of values
of $\sigma $ near $\pi .$ Consequently, no matter how small $\Gamma ,$ $R$
(the curvature radius of the surface of collision $H^3$) is typically of
order $H_{dS}^{-1},$ the curvature radius of the de Sitter space into which the
bubble expands. To obtain a value of $\Omega _0$ close enough to one today, 
 $H_{dS}^{-1}$ must be sufficiently large. As discussed in ref.~\cite{bb},
estimating $R_{min}$ depends on the unknown details of the equation of state 
on the brane arising from the collision, although in a rather 
insensitive way. For purposes
of illustration, we adopt the assumption of a radiation equation of state
in the aftermath of the collision,
obtaining\footnote{Note that the directions of the 
inequalities in eqns.~(66), (69), and (70) of ref.~\cite{bb}
should be reversed.} 
$R\gtorder 10^{24}~\ell ~(m_4\ell )^{-(2/5)}.$
Here $\ell $ is the AdS curvature radius inside the bubble. 
The largest admissible value $\ell \approx 1 ~{\rm mm}$ 
gives $R\gtorder 10^{10} ~{\rm cm}.$
While this large curvature is somewhat of an embarrassment, 
the five-dimensional bulk cosmological constant problem implied here
is orders of magnitude milder than the four-dimensional effective 
cosmological constant today. 
One might hope to evade this restriction by appealing to the anthropic 
principle. In other words, one would have $H_{dS}^{-1}$ small so that
most bubble collisions would result in empty and uninhabitable universes.
We, however, would descend from the collision of one of those rare pairs
for which $\Omega _0$ is not minuscule. Such a scenario, however, would
predict a distribution of $\Omega _0$ seen by us peaked near the lowest
acceptable value, much as in Weinberg's proposed anthropic resolution of
the cosmological constant problem\cite{weinberg}, and this prediction is 
at variance with observation. 

With respect to the second question, we found
that collisions with third bubbles are rare, allowing the surface of
collision to span several curvature radii, a size in excess of that of
the universe observable to us today. Our analysis, however, 
considered only bubbles that collide with the surface of collision. Bubbles
that collide from the side with one of the two bubbles from which our
universe originated do lie in our past lightcone and in principe could affect
us. In the idealised colliding
bubble scenario such a collision simply results in another universe like ours;
however, the possibility that some of the 
debris from the collision escapes into the bubble interior and subsequently
propagates to our universe cannot be ruled out. 
We have not considered such possible collisions. 

\vskip 5pt

\noindent
{\bf Acknowledgements:} We thank J.J. Blanco-Pillado, J. Garriga, A.
Vilenkin, and especially A. Linde for useful discussions. 
CC was supported by the Programa PRAXIS XXI of 
the Funda\c c\~ao para a Ci\^encia e a Tecnologia.
MB would like to thank Mr Denis Avery for
supporting this work through the Stephen W. Hawking Fellowship in 
Mathematical Sciences at Trinity Hall.

\end{document}